\documentclass[sigconf,natbib]{acmart}

\usepackage{amsmath, array, graphicx}
\usepackage{multirow}
\usepackage{booktabs}
\usepackage{tabularx}
\usepackage{subcaption}
\usepackage{balance}
\usepackage[
]{siunitx}
\usepackage{color}
\usepackage{svg}

\definecolor{ballblue}{rgb}{0.13, 0.67, 0.8}
\definecolor{rhodamine}{rgb}{0.8, 0.2, 0.8}

\newcommand{\jaccardindex}{\it Jaccard Index}
\newcommand{\coverageratio}{\it Coverage Ratio}

\newcommand{\ssn}{{\phantom{\dagger}}}
\newcommand{\ssa}{\dag}
\newcommand{\ssb}{\ddag}

\copyrightyear{2023} 
\acmYear{2023} 
\setcopyright{acmlicensed}\acmConference[SIGIR '23]{Proceedings of the 46th International ACM SIGIR Conference on Research and Development in Information Retrieval}{July 23--27, 2023}{Taipei, Taiwan}
\acmBooktitle{Proceedings of the 46th International ACM SIGIR Conference on Research and Development in Information Retrieval (SIGIR '23), July 23--27, 2023, Taipei, Taiwan}
\acmPrice{15.00}
\acmDOI{10.1145/3539618.3591960}
\acmISBN{978-1-4503-9408-6/23/07}

\begin{document}

\settopmatter{
  printacmref=true,
}

\title{Can Generative LLMs Create Query Variants for Test Collections?}
\subtitle{An Exploratory Study}

\author{Marwah Alaofi}
\orcid{https://orcid.org/0000-0002-0008-8650}
\affiliation{
    \institution{RMIT University}
    \city{Melbourne}
    \country{Australia}
}
\email{marwah.alaofi@student.rmit.edu.au}

\author{Luke Gallagher}
\orcid{https://orcid.org/0000-0002-3241-7615}
\affiliation{
\institution{RMIT University}
\city{Melbourne}
\country{Australia}
}
\email{luke.gallagher@rmit.edu.au}

\author{Mark Sanderson}
\orcid{https://orcid.org/0000-0003-0487-9609}
\affiliation{
\institution{RMIT University}
\city{Melbourne}
\country{Australia}
}
\email{mark.sanderson@rmit.edu.au}

\author{Falk Scholer}
\orcid{https://orcid.org/0000-0001-9094-0810}
\affiliation{
\institution{RMIT University}
\city{Melbourne}
\country{Australia}
}
\email{falk.scholer@rmit.edu.au}

\author{Paul Thomas}
\orcid{https://orcid.org/0000-0003-2425-3136}
\affiliation{
\institution{Microsoft}
\city{Adelaide}
\country{Australia}
}
\email{pathom@microsoft.com}

\begin{abstract}
This paper explores the utility of a Large Language Model (LLM) to automatically generate queries and query variants from a description of an information need.
Given a set of information needs described as backstories, we explore how similar the queries generated by the LLM are to those generated by humans. We quantify the similarity using different metrics and examine how the use of each set would contribute to document pooling when building test collections. Our results show potential in using LLMs to generate query variants. While they may not fully capture the wide variety of human-generated variants, they generate similar sets of relevant documents, reaching up to 71.1\% overlap at a pool depth of 100.

\end{abstract}



\maketitle

\section{Introduction and Background}
Information Retrieval (IR) has been dedicated to delivering relevant information
in response to user queries. The realization of this objective has been
facilitated by the use of offline test collections, which often provide a single
representation (query) for each information need.
The single query assumption is convenient for numerous reasons.
It helps to make the judging process economically viable (along with the system
pooling approach in the Cranfield paradigm of test collection
construction~{\cite{Voorhees2002ThePhilosophy}}), and it has provided a
consistent, reusable environment for the development of
retrieval systems and evaluation measures.
The importance of query variations for enumerating relevant documents in a test
collection dates back several decades~{\cite{Jones1977Report}} and previous tracks
at TREC have explored the significance of such variance~{\cite{Buckley1999TheTrec8}}.
More recently, there has been a line of research providing further insights from
the user perspective with the advent of crowd-sourcing
technologies~\cite{Bailey2016Uqv100,Moffat2017Incorporating}.

Query variants are alternative formulations of the same information need. For example, \textit{``what hiking options are there in summer in sangre de cristo''} and \textit{``sangre de Cristo, new mexico hiking''} are both query variants generated in response to the same information need, i.e., finding information for a hiking trip in the Sangre de Cristo mountain region during summer. \citet{Bailey2016Uqv100} showed that given the same backstory, users generate about 57 query variants on average - which is anticipated to increase as the number of participants grows.
Similar findings are reported by \citet{Mackenzie2020Ccnewsen} for queries generated to find additional information in response to document summaries. 

The impact of query variants on retrieval has been empirically demonstrated in prior research. \citet{Culpepper2021Topic} showed that variants impact effectiveness substantially more than that due to topic or ranking models.   \citet{Penha2022Evaluating} tested the impact of variants using neural and transformer-based answer retrieval models. Their experimental results demonstrated a 20\% effectiveness drop on average.

\citet{Alaofi2022Where} empirically demonstrated the impact of query variants on a commercial search engine and different inverted indexes. Their results point to a substantial retrieval inconsistency and a concerning impact of variants on document retrievability. 

Query variants have also been demonstrated to have a comparable impact on the pool size as that of systems, calling to consider incorporating them when building test collections \cite{Moffat2015Pooled}. The current abstraction of one query per topic in the
majority of test collections raises two concerns about how realistic system
evaluations are: (1) is limiting system evaluation to a single representation of
the information need appropriate, and; (2) how the test collection is constructed in
the first place. Can we offer a solution to (1) through the use of the LLMs to
generate human-like queries? Or perhaps have them at least act in place to
generate similar pools for constructing test collections (2)? 

Query variants have primarily been obtained through crowd-sourcing \cite{Mackenzie2020Ccnewsen,Bailey2016Uqv100}, a process that is expensive to scale and may not be an accurate representation of query variants as the information needs are not naturally derived. A study by \citet{Zhang2019Generic}, uses click graphs to collect query variants, based on the assumption that queries leading to the same click originate from the same information need. It is not clear if this assumption always holds true as a shared click may not necessarily indicate a shared intent and many shared intents may not lead to a shared click.
This requires external labeling which is difficult to achieve objectively, and as in the case of crowd-sourcing is expensive to scale. 

In-Context Learning (ICL) \cite{Brown2020Language} emerges as a promising Natural Language Processing paradigm where no large domain-specific datasets are required to fine-tune LLMs on a specific downstream task. Instead, the LLMs are conditioned using a `context' which is simply a textual description of the task with a few or even zero examples - often referred to as a few-, one-, or zero-shot learning. This approach achieves promising results and has surpassed some of the state-of-the-art models in some tasks. This holds the promise of addressing the challenge posed by the scarcity of large query variant datasets. 

ICL has been recently used in IR, mainly to generate synthetic queries given
documents \cite{Bonifacio2022Inpars,Jeronymo2023Inparsv2,Dai2022Promptagator}. The synthetic query-document
pairs are then used to train a retrieval model. This approach builds on earlier
efforts which, prior to indexing, used fine-tuned LLMs to extend documents by
generating relevant queries, an approach that was simple yet effective to
surpass state-of-the-art models on retrieval benchmarks
\cite{Nogueira2019Document}. Though the advances this research direction has
made, it specifically aims to harness the power of LLMs to boost effectiveness
scores by using some `representative' queries, which may undergo some automated
quality/consistency filtering \cite{Dai2022Promptagator,Gospodinov2023Doc2query},
resulting in a query set that improves performance but may not necessarily
represent users. 

The aim of this study is to explore using an LLM (GPT-3.5) as an alternative method to generate query variants given backstories (i.e., information need statements). We aim to compare the LLM-generated queries to human-generated ones. We approach this by quantifying the direct similarity between the two sets of queries and examining how they behave when used for pool construction. In particular, we pose the following questions:
\begin{itemize}
   \item [\textbf{RQ1}] Can an LLM, with one-shot learning, generate queries that are similar or perhaps identical to the ones generated by humans? 
   \item [\textbf{RQ2}] How do both sets compare when used for document pooling when constructing test collections?  
\end{itemize}

\section{Experiment Design}
We detail the query sets and metrics used in the experiment.
\subsection{Query Sets, Model Prompting, and Runs}
We use two sets of query variants: human and GPT-3.5 generated.

The human-generated query variants -- referred to as the \textit{human set} --
were collected via crowd-sourcing as part of the UQV100 test collection \cite{Bailey2016Uqv100}, which has one hundred backstories to describe one hundred search topics derived from the TREC 2013 and 2014 web track. Crowd workers were asked to read a backstory and formulate an initial query for the search task.

To generate the \textit{GPT sets}, we use the same backstories from UQV100 to prompt the model. We use the \textit{text-davinci-003} model.\footnote{Last accessed on 2 February 2023} 
This model is trained similarly to  InstructGPT \cite{Ouyang2022Training} using reinforcement learning with reward models fine-tuned on human preferences.\footnote{https://platform.openai.com/docs/model-index-for-researchers} We experiment with different temperature settings $temp = \{0.0, 0.5, 1.0\}$, a parameter that controls how deterministic the model is in generating the text.

\begin{figure}

    \centering
    \setlength{\abovecaptionskip}{1.5pt}
    \setlength{\belowcaptionskip}{1.5pt}
    \includegraphics[width=\columnwidth]{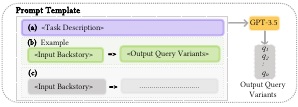}
    \caption{The prompt used to feed the GPT-3.5 model.  
    }
    \label{fig-prompt}
\end{figure}

We prompt the model using the template in Figure \ref{fig-prompt}. The prompt has (a) a \textit{task description}, (b) an \textit{example}, and (c) an \textit{input backstory} for the model to generate the corresponding query variants. The task description is a natural language specification of the task, which provides some context and details to guide the model toward the expected distribution of the queries per backstory and the average number of words per query, with specific values of these settings based on prior research in query variant analysis \cite{Bailey2016Uqv100,Mackenzie2020Ccnewsen}. 
\begin{figure}
  \centering
    \setlength{\abovecaptionskip}{1.5pt}
    \setlength{\belowcaptionskip}{1.5pt}
    \includegraphics[width=\columnwidth]{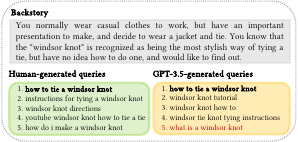}
    \caption{Randomly selected example variants generated by humans and GPT-3.5 ($temp=0.5$). The variant in bold is reproduced by GPT-3.5 and the one in red appears less appropriate, although similar ones are generated by humans, e.g., ``windsor knot wiki''.
    }
    \label{fig-example-variants}
\end{figure} 

We follow a one-shot learning approach, in which the prompt contains an example input backstory with its associated human-generated output queries, randomly selected from UQV100. The same random example (i.e., topic $275$) is used to prompt the model to generate query variants for the remaining 99 backstories. In order to avoid the influence of observed data, topic $275$ has been excluded from our analysis.

We investigated zero-shot learning with no examples provided to the model. However, this approach resulted in the model producing long variants that closely resembled natural language questions. Few-shot learning, where multiple examples are provided to the model, might have produced better results \cite{Brown2020Language}, but as we were limited by the number of available backstories we opted to use one-shot learning. The distribution of all query sets are presented in Table \ref{variants-stats}. Some example query variants from the human set and one of the GPT sets for a given backstory are shown in Figure \ref{fig-example-variants}.

We generate runs for the human set and the GPT sets using Anserini \cite{Yang2017Anserini} BM25 ($b=0.4$, $k1=0.9$) on the ClueWeb12-B corpus\footnote{https://www.lemurproject.org/clueweb12.php/} - which is also the corpus that was used to create relevance judgments for the UQV100 test collection. 
The prompt template we use and the generated GPT sets are publicly
available to aid
reproducibility.\footnote{\href{https://github.com/MarwahAlaofi/SIGIR-23-SRP-UQV100-GPT-Query-Variants}{https://github.com/MarwahAlaofi/SIGIR-23-SRP-UQV100-GPT-Query-Variants}}

\begin{table}
\caption{Query (Q) statistics of the human set and the three GPT sets under different temperature $temp$ settings.
}
\centering
\setlength{\tabcolsep}{3pt}
\begin{tabular}{@{}lrrrrr@{\hspace{2pt}}r@{}}
\toprule
\multirow{2}{*}{\textbf{\shortstack{Query \\Variant Set}}}& \multicolumn{5}{c}{\textbf{Number of Variants}} & \multirow{2}{*}{\textbf{\shortstack{Avg.\\Words/Q}}}\\
\cmidrule(lr){2-6}
& \multicolumn{1}{c}{\textbf{Total}} & \multicolumn{1}{c}{\textbf{Unique}} & \multicolumn{1}{c}{\textbf{Min.}} & \multicolumn{1}{c}{\textbf{Max.}} & \multicolumn{1}{c}{\textbf{Avg.}} & \\
\midrule
Human & 10726 & 5681 & 19 & 101 & 57.38 & 5.34 \\
GPT ($temp = 0.0$) & 4803 & 3638 & 11 & 172 & 36.75 & 5.95 \\
GPT ($temp = 0.5$) & 3061 & 2999 & 12 & 88 & 30.29 & 4.86 \\
GPT ($temp = 1.0$) & 2725 & 2719 & 12 & 48 & 27.46 & 4.65 \\
\bottomrule
\end{tabular}
\label{variants-stats}
\end{table}

\subsection{Metrics for Query Similarity}
In addressing \textbf{RQ1}, we assume that the human set is the ideal set of
query variants and measure how similar the GPT sets are to that set. We quantify that by measuring the average {\jaccardindex} between the human set and the three GPT sets in which the overlapping queries are an exact match between the two sets. 

As keyword-based ranking models treat queries with slight variations equally, we
incrementally relax the matching condition using text transformations over the two
sets and report the average {\jaccardindex} score using the unique queries generated after each transformation. 
Specifically, the overlap between the two sets is quantified by initially determining the exact match of raw queries in both sets. This matching condition is then relaxed to cumulatively allow for variations in punctuation (T1), word forms (T2), stop words (T3), and word order (T4). 
We also show the {\coverageratio}, which quantifies the proportion of queries from the human set that are successfully reproduced in the GPT sets using the aforementioned matching conditions.

\subsection{Metrics for Retrieval Similarity}
To address \textbf{RQ2}, we find the overlap between the union of documents returned
from the variants of the human set and the GPT sets. This
measure quantifies the similarity of the two sets in their utility for constructing document pools. The overlap is quantified using the {\jaccardindex}, measured at different depths. While we are interested in measuring the overall overlap in documents (regardless of their relevance judgments), we believe that the impact of the overlap of relevant documents in particular holds greater importance in regards to system evaluation when constructing test collections. That is, if a GPT set fails to retrieve irrelevant documents that are retrieved by the human set, those documents will remain unjudged and thus treated as irrelevant in most effectiveness metrics, unless using metrics that account for uncertainty (e.g., RBP).

We use the relevance judgments provided with the UQV100 collection and measure the overlap by considering relevant documents alone, i.e., those rated as \textit{Essential}; \textit{Very Useful}; \textit{Mostly Useful} or \textit{Slightly Useful}.   

Different properties of the document pool are computed, mainly the \textit{pool size growth} following \cite{Moffat2015Pooled}, to measure the diversity of the GPT sets in comparison to the human set. We hypothesize that a diverse set of query variants given a topic is likely to retrieve different documents leading to a higher growth rate than a set of similar queries. This diversity is also examined using \textit{Rank-Biased Overlap (RBO)} \cite{Webber2010ASimilarity} to quantify the consistency between the retrieved documents of the query variants given a topic. A topic RBO score is the average score over all topic-variant pairs.
Different query effectiveness metrics are also computed for comparison.

\section{Results and Discussion}
We examine query and retrieval similarity.

\begin{figure}[t]
    \setlength{\abovecaptionskip}{0pt}
    \setlength{\belowcaptionskip}{0pt}

    \centering
    \begin{subfigure}{0.48\columnwidth}
        \includegraphics[width=\linewidth]{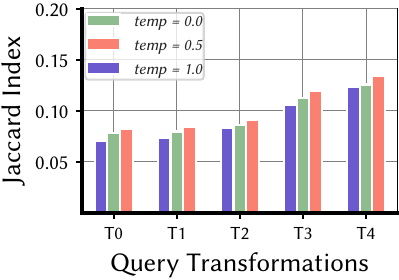}
        \label{labela}
    \end{subfigure}
    \hfill
    \begin{subfigure}{0.48\columnwidth}
        \includegraphics[width=\linewidth]{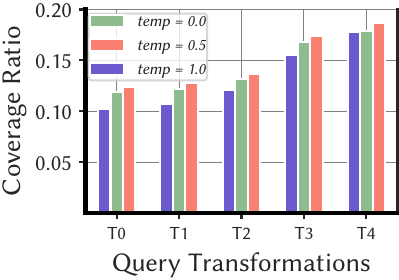}
        \label{labelb}
    \end{subfigure}
    \caption{Jaccard index (left) and coverage ratio (right) between the human and the GPT sets under different temperature $temp$ settings. \textit{T0 denotes the original query set}.
    }
    \label{fig-query-sets}
\end{figure}

\subsection{Query Sets Similarity}

Figure \ref{fig-query-sets} shows the overlap between the human set and the GPT sets as measured by the Jaccard index and coverage ratio under four matching conditions. Results indicate a minimum of 7.1\% Jaccard index between the GPT sets and the human set, with GPT sets demonstrating exact coverage of at least 10.3\% of the human-generated queries. As expected, the queries demonstrate a greater degree of overlap as they undergo successive transformations, ultimately reaching a maximum Jaccard index of 13.5\% and coverage ratio of 18.7\% when using a temperature of 0.5.

\begin{table*}[h!]
   \caption{
    Average effectiveness metrics, RBO and pool properties at depth 10 for the
    human set and the GPT sets given all variants across all topics. RBP and RBO are 
    measured using $p=0.9$. 
    Entries annotated with $\ssa$ and $\ssb$ respectively indicate statistical
    significance for a Bonferroni pairwise
    $t$-test at $p<0.05$ and $p<0.01$ compared to the human query set baseline. Topic 275 (the example used to prompt the model) was removed from the computation and results are replicated independently. 
    }
  \centering
  \newcommand{\MCC}[1]{\multicolumn{1}{c}{#1}}
  \begin{tabular}{lccclrcc}
    \toprule
    \multirow{2}{*}{\textbf{Variant Set}}
      & \multirow{2}{*}{\textbf{P@10}}
      & \multirow{2}{*}{\textbf{NDCG@10}}
      & \multirow{2}{*}{\textbf{RBP}}
      & \multirow{2}{*}{\textbf{RBO}}
      & \multicolumn{3}{c}{\textbf{Pool Properties}}
      \\
    \cmidrule(rrr){6-8}
      &
      &
      &
      &
      & \MCC{\textbf{Size}}
      & \textbf{Relevant}
      & \textbf{Unjudged}
      \\

    \midrule
    Human set
      & 0.443$^\ssn$
      & 0.274$^\ssn$
      & 0.406 +0.111$^\ssn$
      & 0.201
      & 190.69
      & 0.30
      & 0.13
      \\
    GPT ($temp = 0.0$) 
      & 0.386$^\ssb$ 
      & 0.246$^\ssa$ 
      & 0.358 +0.254$^\ssb$
      & 0.235$^\ssa$ 
      & 94.42
      & 0.29
      & 0.33
      \\
    GPT ($temp = 0.5$) 
      & 0.393$^\ssb$
      & 0.249$^\ssa$
      & 0.360 +0.238$^\ssb$
      & 0.220
      & 93.55
      & 0.29
      & 0.31
      \\
    GPT ($temp = 1.0$) 
      & 0.384$^\ssb$
      & 0.240$^\ssb$
      & 0.355 +0.263$^\ssb$
      & 0.235$^\ssa$ 
      & 105.21
      & 0.27
      & 0.37
      \\
    \bottomrule
    \end{tabular}
    \label{tbl-results}
\end{table*}

While the observed overlap does not seem to indicate a high similarity, they should be interpreted within the limitation of the UQV100 human set - which is still somewhat artificial. That is, it cannot be conclusively stated that the unique query variants generated by GPT cannot be written by humans should we have more participants. The use of the UQV100 human-generated queries as a reference point to an ideal set of variants is not realistic, and while this comparison helps understand the capability of GPT, it may limit our interpretation as an exhaustive set of query variants given a topic would never exist. 
Incorporating human evaluation to assess the extent to which the GPT sets approximate human queries may yield more precise conclusions. This is, however, a question to be explored in future research.

\subsection{Retrieval Similarity}

\begin{figure}[]
    \setlength{\abovecaptionskip}{0pt}
    \setlength{\belowcaptionskip}{0pt}

    \centering
    \begin{subfigure}{0.48\columnwidth}
        \includegraphics[width=\linewidth]{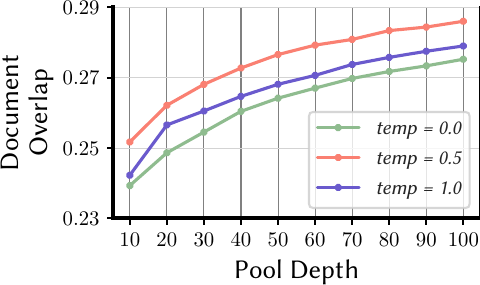}
        \label{labela}
    \end{subfigure}
    \hfill
    \begin{subfigure}{0.48\columnwidth}
        \includegraphics[width=\linewidth]{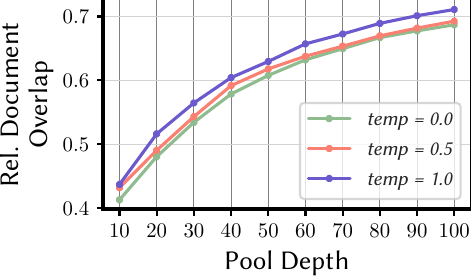}
        \label{labelb}
    \end{subfigure}
    \caption{The average Jaccard index between the documents retrieved by the human set and the GPT sets at different depths given all documents (left) and relevant documents only (right).
    }
    \label{fig-doc-overlap}
\end{figure}

Figure \ref{fig-doc-overlap} shows the overlap between the document pools generated in response to the human set and the GPT sets at different depths measured by the Jaccard index.
A relatively high average overlap of document pools is observed between the GPT sets and the human set, which increases as we increase the depth. When considering relevant documents only, the overlap is considerably high. With a temperature of 1.0, for example, the pools overlap at 43.7\% at depth 10. This increases to 71.1\% when examining the pools at depth 100.
\begin{figure}[]
  \centering
    \includegraphics[width=0.60\columnwidth]{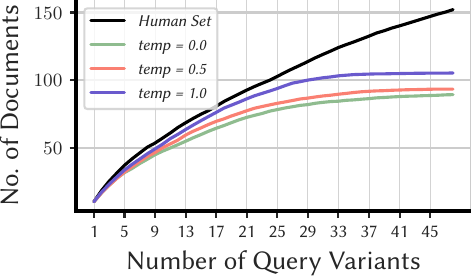}
    \caption{The average pool size at depth 10 as more variants are added. The growth lines are cut at 48, which is the maximum number of variants generated by GPT at $temp = 1$. }
    \label{fig-doc-growth}
\end{figure}

Effectiveness metrics, RBO scores and pool properties are given in Table \ref{tbl-results}.
It is evident that the human-generated variants yield a larger pool size, almost
double that generated by any GPT sets, and which also grows faster (see Figure \ref{fig-doc-growth}). This indicates a possible
higher diversity (e.g., more distinct query terms) in the human set. This
observation is supported by a lower consistency of the variants from the
human set, as measured by RBO (significant with the temperature set to 0.0 or 1.0). 

Human-generated variants are significantly more effective across all metrics. GPT queries, on the other hand, have higher residuals which indicate that, given more judgments, they may achieve higher effectiveness scores. This is also reflected in the higher proportion of the unjudged documents returned in the GPT generated pools. It would be interesting to further investigate the unjudged portion of the GPT sets to understand whether they retrieve relevant documents that were not found through the human set.

\section{Conclusions and Future Work}
In this paper, we posed the following questions:
\begin{itemize}
   \item [\textbf{RQ1}] Can an LLM, with one-shot learning, generate queries that are similar or perhaps identical to the ones generated by humans? 
   \item [\textbf{RQ2}] How do both sets compare when used for document pooling when constructing test collections? 
\end{itemize}

We found that for {\bf RQ1}, GPT reproduced a reasonable portion of the human-generated queries. The similarity to human queries is yet to be fully understood given the limitation of the human set.

For {\bf RQ2}, we found that GPT queries seem to have a substantial overlap in the pool of documents, particularly when we consider the relevant set alone. At 71.1\% overlap at depth 100, GPT shows potential for replacing human query variants with synthetically generated ones during document pool construction.

This work presents a new opportunity to conveniently expand existing test collections, particularly those resembling TREC, which have information need statements that can be employed to condition LLMs. 
Further research could explore advanced prompting techniques and compare our approach of using an LLM to generate query variants from backstories with previous query simulation methods (e.g., \cite{Jordan2006Using, Breuer2022Validating}), which were used to generate query variants from source documents.

\begin{acks}
  Marwah Alaofi is supported by a scholarship from Taibah University, Saudi Arabia.
  This work was also supported by the \grantsponsor{ARC}{Australian Research
  Council}{https://www.arc.gov.au/} (\grantnum{ARC}{DP180102687}, \grantnum{ARC}{CE200100005}).
  We thank the anonymous reviewers for their helpful feedback.
\end{acks}

\bibliographystyle{ACM-Reference-Format}
\balance
\bibliography{references}
\end{document}